\documentstyle[11pt,aaspp4,psfig]{article}

\def\degree{{$^\circ$}}

\def\puncspace{\ifmmode\,\else{\ifcat.\C{\if.\C\else%
\if,\C\else\if?\C\else\if:\C\else\if;\C\else\if-\C\else%
\if)\C\else\if/\C\else\if]\C\else\if'\C%
\else\space\fi\fi\fi\fi\fi\fi\fi\fi\fi\fi}%
\else\if\empty\C\else\if\space\C\else\space\fi\fi\fi}\fi}%
\def\SP{\let\\=\empty\futurelet\C\puncspace}

\def\ee#1{\ifmmode {} \times 10^{#1} \else ${} \times 10^{#1}$\fi}
\def\sub#1{\ifmmode _{#1} \else $_{#1}$\fi}
\def\sup#1{\ifmmode ^{#1} \else $^{#1}$\fi}

\def\about{\ifmmode \sim \else {$\sim\,$}\fi}
\def\lta{\ifmmode {\,\mathbin{\lower 3pt\hbox   
    {$\,\rlap{\raise 5pt\hbox{$\char'074$}}\mathchar"7218\,$}}}
    \else {${\mathbin{\lower 3pt\hbox
    {$\rlap{\raise 5pt\hbox{$\char'074$}}\mathchar"7218\,$}}}
    $}\fi}
\def\gta{\ifmmode {\mathbin{\lower 3pt\hbox   
    {$\,\rlap{\raise 5pt\hbox{$\char'076$}}\mathchar"7218\,$}}}
    \else {${\mathbin{\lower 3pt\hbox
    {$\rlap{\raise 5pt\hbox{$\char'076$}}\mathchar"7218\,$}}}
    $}\fi}

 \mathcode`*="002A   

\def\degree{{\ifmmode ^\circ \else $^\circ$\fi}}

\def\km{{\hbox{km}}\SP}

\def\fu#1{\leavevmode\hbox{4U~#1}\SP}

\def\ks#1{\leavevmode\hbox{KS~#1}\SP}

\def\psr#1{\leavevmode\hbox{PSR~#1}\SP}

\def\mdot{{\ifmmode \dot M \else {$\dot M$}\fi}}
\def\mdote{{\ifmmode \dot M_E \else {$\dot M_E$}\fi}}
\def\mdoti{{\ifmmode \dot M_i \else {$\dot M_i$}\fi}}
\def\msun{{\ifmmode M_\odot \else {$M_{\odot}$}\fi}}

\begin{document}

\lefthead{Miller \& Lamb}
\righthead{Bounds on Neutron Stars from Oscillations During X-ray Bursts}

\title{Bounds on the Compactness of Neutron Stars from
Brightness Oscillations\\
 During X-ray Bursts}

\author{M.\ Coleman Miller}
\affil{Department of Astronomy and Astrophysics, University of Chicago\\
       5640 South Ellis Avenue, Chicago, IL 60637, USA\\
       miller@bayes.uchicago.edu}
\authoremail{miller@bayes.uchicago.edu}

\author{Frederick K.\ Lamb}
\affil{Department of Physics and Department of Astronomy\\
       University of Illinois at Urbana-Champaign\\
       1110 W. Green St., Urbana, IL 61801-3080, USA\\
       f-lamb@uiuc.edu}

\begin{abstract}

The discovery of high-amplitude brightness
oscillations at the spin frequency or its first overtone in six
neutron stars in low-mass X-ray binaries during type~1
X-ray bursts provides
a powerful new way to constrain the compactness of these
stars, and hence to constrain the
equation of state of the dense matter in all neutron stars. 
Here we present the
results of general relativistic calculations of the maximum
fractional rms amplitudes that can be observed during bursts.
In particular, we determine the dependence of the amplitude
on the compactness of the star, the angular dependence
of the emission from the surface, the rotational velocity
at the stellar surface, and whether there are one or two
emitting poles. We show that if two poles
are emitting, as is strongly indicated by independent evidence in
\fu{1636$-$536} and \ks{1731$-$26}, the resulting
limits on the compactness of the star can be extremely
restrictive. We also discuss the expected amplitudes of X-ray
color oscillations
and the observational signatures necessary to derive
convincing constraints on neutron star compactness from the
amplitudes of burst oscillations.

\end{abstract}

\keywords{stars: neutron --- equation of state --- gravitation
--- relativity --- X-rays: bursts}

\section{INTRODUCTION}

The determination of the equation of state of neutron stars
has been an important goal of nuclear physics for more than
two decades. Progress toward this goal can be made by
establishing astrophysical constraints as well as by improving
our understanding of nuclear forces.

Many ways of deriving constraints from astrophysics have
been explored. One of the best known is pulse timing of
pulsars in binary systems. 
Although binary pulsar timing has made possible
stringent 
tests of general relativity (see, e.g., Taylor 1992), the 
$\approx 1.4\,M_\odot$ masses derived from timing (see Thorsett
et al.\ 1993)
are allowed by all equations of state based on realistic nuclear
physics, and hence these observations have not eliminated any of 
the equations of state currently being considered.
The highest known neutron
star spin frequency, the 643~Hz frequency of \psr{1937$+$21}, is 
also allowed
by all equations of state currently under consideration. 
Radius estimates based on the energy spectra of type~1 X-ray bursts 
and on observations of thermal emission from the surfaces of neutron stars
are more restrictive
in principle but currently have large systematic uncertainties (see
Lewin, van Paradijs, \& Taam 1993; \"Ogelman 1995).

The discovery of 
high-frequency brightness oscillations from 
sixteen neutron stars in low-mass X-ray binaries with the {\it Rossi}
X-ray Timing Explorer (RXTE) holds great promise for providing
important new constraints.
Oscillations are observed both in the persistent X-ray emission and
during type~1 X-ray bursts. The kilohertz quasi-periodic
oscillations (QPOs)
observed in the persistent emission have
high amplitudes and relatively high coherences (see van der Klis
1997). A pair of kilohertz
QPOs is commonly observed from a given source. Although
the frequencies of these QPOs vary by up to a factor of $\sim$2,
the frequency separation $\Delta\nu$ between a pair of kilohertz
QPOs appears to be constant in almost all cases.
In both the sonic-point (Miller, Lamb, \& Psaltis 1996) and
magnetospheric (Strohmayer et al.\ 1996) beat-frequency 
interpretations, the higher frequency in a pair is the
orbital frequency at the inner edge of the Keplerian flow,
whereas the lower
frequency is the beat of the stellar spin frequency
with this frequency. Such high orbital frequencies
yield interesting bounds on the masses and radii of these
neutron stars and interesting constraints on the equation of
state of neutron star matter (Miller et al.\ 1996).

The brightness oscillations observed during type~1 X-ray 
bursts are different in
character from the QPOs observed in the persistent emission.
Only a single oscillation has been observed from each source during a
type~1 X-ray burst, and the oscillations in the tails of 
bursts appear
to be highly coherent (see, e.g., Smith, Morgan, \& Bradt 1997), 
with frequencies
that are always the same for a given source (comparison of
burst oscillations from \fu{1728$-$34} over about a year shows that
the timescale for any variation in the oscillation frequency is
$\gta 3000$~yr; Strohmayer 1997).
The burst oscillations in \fu{1728$-$34} (Strohmayer
et al.\ 1996; Strohmayer 1997) and \fu{1702$-$42} (Swank 1997)
have frequencies that are consistent
with the separation frequencies of their kilohertz QPO pairs. 
The burst oscillations in \fu{1636$-$536} (Zhang et al.\ 1996) and
\ks{1731$-$260} (Smith et al.\ 1997; Wijnands \& van der Klis
1997), have frequencies that are consistent with twice 
the separation frequencies of their kilohertz QPO pairs. 
The burst oscillations are thought to be caused by emission
from one or two nearly identical emitting poles, producing
oscillations at the stellar spin frequency or its first
overtone, respectively.
If so, the anisotropy observed at infinity
of the radiation emitted
from the surface, and hence the amplitude of the burst oscillations,
typically decreases with increasing gravitational 
light deflection. The burst oscillations can therefore 
be used to constrain the compactness of neutron stars and
the equation of state of neutron star matter (Strohmayer 1997).

In this Letter we present the results of general relativistic
calculations to determine the maximum relative amplitudes of
burst oscillations, as a function of stellar compactness,
following the procedure described by
Pechenick, Ftaclas, \& Cohen (1983; see also
Chen \& Shaham 1989).
We go beyond these previous treatments by using a
more realistic angular intensity distribution at
the stellar surface, by including the angle dependence
of the energy spectrum, by considering the effects of X-ray
color oscillations, and by computing the amplitudes at overtones.
We show that gravitational
light deflection typically reduces the relative amplitude of the
oscillations produced by two poles compared to one pole and that
effects other than Doppler shifts and aberration are second-order
in the rotation rate.
We describe our assumptions and method in \S~2. In \S~3
we present our results and discuss
the observational signatures that allow the 
derivation of robust limits on the compactness of the
burst sources.

\section{ASSUMPTIONS AND METHOD}

\subsection{Assumptions}

The main purpose of our calculations is to derive constraints on the
compactness of the neutron stars that display oscillations during
type~1 X-ray bursts. To do this, we
must determine the largest possible amplitude of an
oscillation for a given compactness. We therefore make the following
assumptions, which maximize the oscillation amplitude (these are 
discussed further in \S~3):

(1)~The radiation that we see comes directly from the stellar
surface.

(2)~The star radiates from one or two emitting poles; 
the rest of the surface is dark.

(3)~The pole or poles are pointlike and located in the rotational equator.

(4)~If there are two poles, they are identical and antipodal.

(5)~The observer's line of sight is in the rotational equator.

The effects on the spacetime of the star's rotation change the
oscillation amplitude only to second order (see
Lamb \& Miller 1995 and Miller \& Lamb 1996 for details),
and are much smaller than first- and second-order Doppler
effects for the relatively slow rotation rates of interest.
We therefore
neglect them, performing all calculations in the Schwarzschild
spacetime.

\subsection{Calculational Method}

We compute the intensity seen by an observer when an
emitting pole is at a given rotational phase using the
procedure discussed by Pechenick et al.\ (1983). 
Suppose that the emitting pole
is at azimuthal angle $\phi=0$ and that the azimuthal angle
of the photon when it reaches infinity is $\phi_\infty$.
The angle $\phi_\infty$ is a maximum when the
photon is emitted tangentially to the stellar surface, and is
then (Pechenick et al.\ 1983, eq. [2.13])
\begin{equation}
\phi_{\rm max}=\int_0^{M/R}\left[(1-{2M\over R})
\left({M\over R}\right)^2-(1-2u)u^2\right]^{-1/2}\,du\; .
\end{equation}
Here and below we use geometrized units in which $G=c\equiv 1$.

Suppose that the observer is at infinity at 
azimuthal angle $\phi_{\rm obs}$. If $\phi_{\rm obs}>\phi_{\rm max}$ 
the observer cannot see the pole. If $\phi_{\rm obs}<\phi_{\rm max}$
we solve for $u_b\equiv M/b$, the
reciprocal of the impact parameter to the pole, using 
(Pechenick et al.\ 1983, eq. [2.12]):
\begin{equation}
\phi_{\rm obs}=\int_0^{M/R}\left[u_b^2-(1-2u)u^2\right]^{-1/2}\,du\; .
\end{equation}
In the Schwarzschild spacetime,
the angle $\psi$ to the normal at the stellar surface of the
light ray that has impact parameter $b$ at infinity is given implicitly by
$\cos\psi=(1-b^2/b_{\rm max}^2)^{1/2}$, where
the maximum impact parameter is $b_{\rm max}=R(1-2M/R)^{-1/2}$
(see, e.g., Abramowicz, Ellis, \& Lanza 1990; Miller \& Lamb 1996).
The flux measured by the observer
is proportional to $I_\nu[\psi(\phi_{\rm obs})]\cos\psi
(\phi_{\rm obs})$, where
$I_\nu(\psi)$ is the specific intensity at photon frequency
$\nu$ and angle $\psi$ from the normal to the pole. As usual, 
$I_\nu/\nu^3$ and $I/\nu^4$ are constant, where
$I$ is the frequency-integrated
specific intensity. Hence, changes in $I_\nu$ and $I$ caused
by Doppler shifts and redshifts
are easily computed by calculating the change
in the frequency of the photons. 

If the star were 
not rotating and the emission from the pole were isotropic, then 
$I(\psi)$ would be constant. However, because the star is rotating,
the intensity distribution is aberrated and 
the photon frequency is Doppler-shifted by
the factor $1/[\gamma(1-v\cos\psi)]$, where 
$v$ is the velocity at the stellar equator and
$\gamma=(1-v^2)^{-1/2}$.
Also, the specific intensity is not expected to be isotropic
even in the rest frame of the stellar surface, because radiation
that diffuses outward through an optically thick scattering
atmosphere emerges with a specific intensity distribution
$I_{\rm sc}(\psi)$ that is
nearly three times as bright along the normal as in directions
tangential to the surface (see Chandrasekhar 1960, chapter 3).
Moreover, the energy spectrum will also, in general, depend
on the angle from the normal. The angle dependence of the
actual spectrum is likely to be complicated (see, e.g.,
Lewin, van Paradijs, \& Taam 1993). However, the qualitative
effects of angle dependence are illustrated by
a simplified emission model in which the spectrum emerging
at angle $\psi$ is a blackbody spectrum with
a temperature $T$ given by $T^4={3\over 4}
T_{\rm eff}^4({2\over 3}\cos\psi+0.7)$, where $T_{\rm eff}$
is the effective temperature; this approximates the angle
dependence of the energy spectrum from a grey atmosphere.

Using this approach, we compute the brightness seen by
an observer at infinity as a function of rotational phase,
for a given specific intensity distribution and rotation rate.
Intensities can be added linearly, so addition of a second pole is
straightforward. We Fourier analyze the resulting light curve 
to determine the oscillation amplitudes at different harmonics 
of the spin frequency.

\section{RESULTS AND DISCUSSION}

The results of our calculations are summarized in Figure~1,
which shows the fractional rms amplitudes of countrate oscillations
in various situations (the amplitudes
of oscillation in the photon number and total energy fluxes are equal for
nonrotating stars but are generally unequal for rotating
stars).
Figure~1a compares the oscillation amplitude
for isotropic emission and for the peaked emission $I_{\rm sc}(\psi)$
from a single pole.
The rms amplitude of
a non-sinusoidal waveform can be greater than unity, as shown.
Figure~1b shows the
fractional rms amplitude at the first overtone under the same
assumptions, for two identical, antipodal emitting poles.
Figure~1c shows the amplitude as
a function of the angle between the observer's line of sight and the
rotation equator for a single emitting pole. Figure~1d 
illustrates the generation of
harmonics by Doppler shifts and aberration.
Figure~1e shows the amplitude as a function of photon energy
for different
assumptions about the surface velocity and the rest-frame 
angle-dependence of the energy spectrum. Figure~1f plots the
amplitude from two emitting poles as detected by instruments
with different bandpasses,
as a function of the effective temperature of the surface.

These results illustrate several points. (1)~The maximum
relative amplitude is significantly greater when one pole is
emitting than when two poles are emitting. (2)~The amplitude
tends to be higher when the neutron star is less compact, i.e.,
when $R/M$ is larger. The exception occurs for isotropic
emission and very compact neutron stars, $R<4\,M$,
in which case gravitational focusing of emission from the
pole on the opposite
side of the star from the observer can increase the amplitude
(this effect was
discussed by Pechenick et al.\ [1983]). (3)~The angular distribution
of the specific intensity makes a large difference. For example,
for $M=1.4\,M_\odot$, $R=10\,\km=4.8\,M$,
two emitting poles, and
$I(\psi)=I_{\rm sc}(\psi)$, the maximum amplitude
is nearly three times greater than 
the maximum amplitude for isotropic emission.
(4)~Stellar rotation affects the {\it bolometric}
amplitude only to second order. However, rotation shifts
the {\it energy spectrum} by an amount
that is first order in the rotational velocity. As a result,
the measured relative countrate amplitude depends strongly on photon
energy if the energy spectrum within the bandpass
of the instrument is steep, and may be large
even for two emitting poles. As Figure~1f shows,
the measured rms amplitude can exceed
$\sim$70\% for two emitting poles. The amplitude will also
depend strongly on the photon energy if, as expected,
the rest-frame energy spectrum depends on the
angle from the normal, and hence the amplitude is likely to
rise steeply 
with increasing photon energy even if Doppler shifts and
aberration are unimportant. (5)~Rotation generates significant power
at overtones of the spin frequency.
(6)~The oscillation amplitude is proportional to the cosine of
the angle $\alpha$ between the line of sight of
the observer and the rotation equator and hence
the observed amplitude is near its maximum value over a large
solid angle: for example, the observed amplitude for
$\alpha$ between $0^\circ$ and
$30^\circ$, which occurs 50\% of the time for
randomly distributed observers, is within 5\% of the maximum.

These results have been computed assuming that
the only emission from the surface comes from pointlike poles,
that if there are two poles then they are identical and
antipodal, and that the
emitting pole or poles are in the spin equator. These assumptions
produce the largest possible amplitude. Small deviations
from them affect the amplitude of the oscillation only
to second order in the deviation,
as is also evident from the results of Pechenick et al.\ (1983). 
For example, if emission at the pole comes from a cap of
angular radius $\delta\ll 1$ instead of from a point, then
the amplitude is decreased by an amount $\sim\delta^2$.
If there are two
poles and they are offset from antipodal by a small angle $\beta$,
then the amplitude at twice the spin frequency drops by
$\sim\beta^2$. If the axis of the pole or poles is offset from
the spin equator by a small angle $\epsilon$, the amplitude is
diminished by $\sim\epsilon^2$.

If the burst is near its onset or its end,
gas will be accreting onto the neutron star and is
likely to Comptonize the radiation coming from the stellar
surface (see Psaltis, Lamb, \& Miller 1995).
This introduces two competing effects. First, scattering tends
to isotropize the radiation, diminishing the observed 
oscillation amplitude
(see, e.g., Brainerd \& Lamb 1987; Kylafis \& Phinney 1989;
Miller 1997).
Second, relatively small changes in the optical depth or temperature
of the accreting gas with azimuth can change the shape of the spectrum,
producing large changes in the flux at high
energies with azimuth, thereby increasing the observed
amplitude (see Miller, Lamb, \& Psaltis 1997; Miller \& Lamb 1992).
In principle, either effect could dominate.

The bounds on stellar compactness that we have computed
apply only when we are
seeing emission coming directly from the stellar surface.
Strong evidence that we {\it are} observing such emission is
an energy spectrum that is close to
a modified blackbody spectrum, because if the radiation from the
surface undergoes significant Comptonization by the 
accreting gas, the spectrum generally
has an extended high-energy tail
(see, e.g., Psaltis, Lamb, \& Miller 1995). Thus, if a 
modified blackbody spectrum
is observed, the amplitudes of oscillations during X-ray
bursts may be used to derive bounds on the compactness
of the star.

\acknowledgements

We are grateful to Dimitrios Psaltis for many helpful discussions of
the amplitudes of burst oscillations, especially about the role
of X-ray color oscillations. We also thank Tod Strohmayer and Jean Swank
for discussing their burst observations in advance
of publication. This work was supported in part by NSF grants
AST~93-15133 and AST~96-18524 and NASA grant
NAG~5-2925 at the University of Illinois, and NASA grant
NAG~5-2868 at the University of Chicago.

\newpage
\figcaption{
Fractional rms amplitude of countrate oscillations for various
situations.
The observer's line of sight is assumed
to be in the rotational equator except in panel~(c).
Aberration and the Doppler shifts caused by rotation are neglected
in panels (a)--(c). (a)~Amplitude as a function of
neutron star radius for isotropic emission
(dotted line) or the peaked emission
$I_{\rm sc}(\psi)$ (solid line) from a single pole.
(b)~Amplitude as a function of radius
for the same intensity distributions as in part (a),
but for two identical, antipodal emitting poles.
(c)~Amplitude as a function of the angle between the observer's
line of sight and the
rotation equator, for a single emitting pole,
$R=5\,M$, and $I(\psi)=I_{\rm sc}(\psi)$. 
(d)~Amplitude as a function of surface velocity,
for one emitting pole, $R=5\,M$, and $I(\psi)=I_{\rm sc}(\psi)$,
at the fundamental of the spin frequency (solid line), the first
overtone (dashed line), and the second overtone (dotted line).
(e)~Amplitude as a function of photon
energy for one emitting pole (upper three curves) and two
emitting poles (lower three curves), assuming $R=5\,M$ and
$I(\psi)=I_{\rm sc}(\psi)$, for a surface velocity $v=0.1$
and an energy spectrum that is angle-independent in the rest
frame of the stellar surface (solid lines); for $v=0$ and
the angle-dependent spectrum described in the text (dotted
lines); and for $v=0.1$ and the angle-dependent spectrum
(dashed lines). (f)~Amplitude as a function of effective temperature, for
two emitting poles, $R=5\,M$, $I(\psi)=I_{\rm sc}(\psi)$, $v=0.1$, and
the angle-dependent energy spectrum.
The three curves show the amplitude measured by a bolometer
(dotted line) and by instruments
with two different boxcar energy responses:
4--20~keV (solid line) and 2--20~keV (dashed line).
Each detected photon is assumed to produce exactly one count.}

\newpage

\begin{figure*}[t]
\hbox{\hskip 0.01truein
\psfig{file=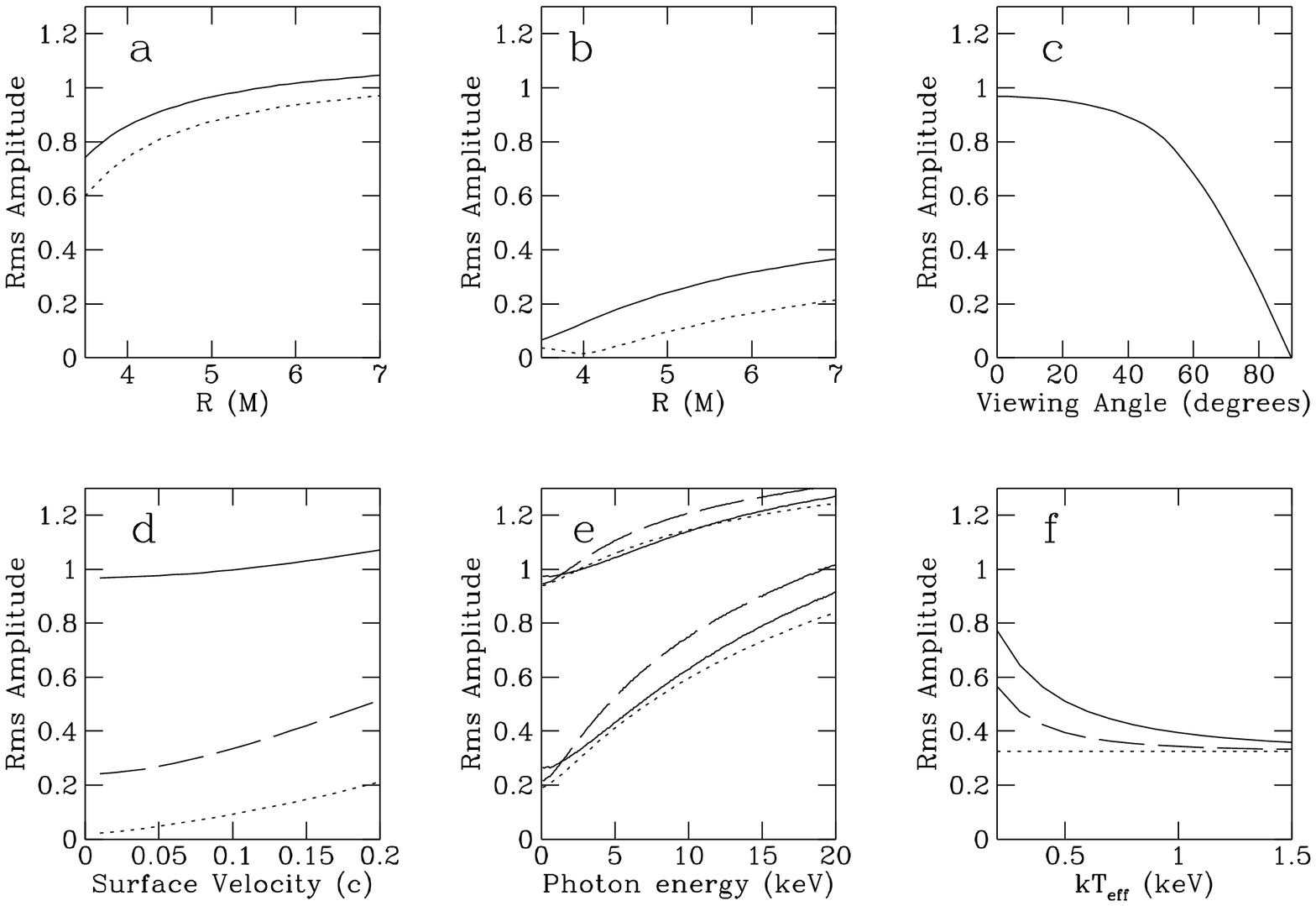,height=6.0truein,width=7.0truein}}
\end{figure*}

\end{document}